\documentclass[11pt,a4paper]{emulateapj}
\slugcomment{ApJ, 771, L2 (2013)}

\usepackage{natbib}

\newcommand{\teff}{${T}_{\mathrm{eff}}$}
\newcommand{\logg}{$\log{g}$}
\newcommand{\msun}{${M}_{\odot}$}

\newcommand{\muhz}{$\mu$Hz}

\shorttitle{Discovery of an Ultramassive Pulsating White Dwarf}
\shortauthors{Hermes et al.}

\begin{document}

\title{DISCOVERY OF AN ULTRAMASSIVE PULSATING WHITE DWARF}
\author{J. J. Hermes\altaffilmark{1,2}, S. O. Kepler\altaffilmark{3}, Barbara G. Castanheira\altaffilmark{1,2}, A. Gianninas\altaffilmark{4}, D. E. Winget\altaffilmark{1,2}, \\ M. H. Montgomery\altaffilmark{1,2}, Warren R. Brown\altaffilmark{5}, and Samuel T. Harrold\altaffilmark{1,2}}

\altaffiltext{1}{Department of Astronomy, University of Texas at Austin, Austin, TX\,78712, USA}
\altaffiltext{2}{McDonald Observatory, Fort Davis, TX\,79734, USA}
\altaffiltext{3}{Instituto de F\'{\i}sica, Universidade Federal do Rio Grande do Sul, Porto Alegre, RS, Brazil}
\altaffiltext{4}{Homer L. Dodge Department of Physics and Astronomy, University of Oklahoma, 440 W. Brooks Street, Norman, OK\,73019, USA}
\altaffiltext{5}{Smithsonian Astrophysical Observatory, 60 Garden Street, Cambridge, MA\,02138, USA}

\email{jjhermes@astro.as.utexas.edu}


\begin{abstract}

We announce the discovery of the most massive pulsating hydrogen-atmosphere white dwarf (WD) ever discovered, GD~518. Model atmosphere fits to the optical spectrum of this star show it is a $12{,}030 \pm 210$ K WD with a \logg\ $=9.08 \pm 0.06$, which corresponds to a mass of $1.20 \pm 0.03$ \msun. Stellar evolution models indicate that the progenitor of such a high-mass WD endured a stable carbon-burning phase, producing an oxygen-neon-core WD. The discovery of pulsations in GD~518 thus offers the first opportunity to probe the interior of a WD with a possible oxygen-neon core. Such a massive WD should also be significantly crystallized at this temperature. The star exhibits multi-periodic luminosity variations at timescales ranging from roughly $425-595$ s and amplitudes up to 0.7\%, consistent in period and amplitude with the observed variability of typical ZZ Ceti stars, which exhibit non-radial $g$-mode pulsations driven by a hydrogen partial ionization zone. Successfully unraveling both the total mass and core composition of GD~518 provides a unique opportunity to investigate intermediate-mass stellar evolution, and can possibly place an upper limit to the mass of a carbon-oxygen-core WD, which in turn constrains Type Ia supernovae progenitor systems.

\end{abstract}

\keywords{stars: individual (GD~518)--stars: white dwarfs--stars: oscillations (including pulsations)--stars: variables: general--stars: evolution--stars: supernovae: general}


\section{Introduction}

White dwarf (WD) stars are stellar remnants composed almost entirely of the inert byproducts of previous nuclear reactions; they are the burnt-out cores of stars with initial masses below about $8.0\pm1.5$ \msun\ \citep{Smartt09,Williams09}.

The majority of WDs have an overall mass near $\sim$0.6 \msun\ \citep{Falcon10,Tremblay11,Kleinman13}. WDs near this canonical mass are expected to harbor remnant carbon-oxygen (CO) cores after core hydrogen burning and subsequent helium burning. However, an isolated progenitor star with an initial mass larger than about $7$ \msun\ will reach sufficiently high temperature to achieve stable carbon burning, and may possibly end up as an ultramassive WD with an oxygen-neon (ONe) or oxygen-neon-magnesium (ONeMg) core, if the progenitor had insufficient conditions to start further nuclear burning and detonate as a Type II supernova \citep{Nomoto84}.

\citet{GBRI97} found that a 9 \msun\ progenitor model undergoes repeated carbon-burning shell flashes when its core exceeds $\sim$1.05 \msun, ultimately ending up as a WD with an ONe core, although rotation may also play a role in the outcome of an intermediate-mass progenitor \citep{Dominguez96}. We note that there are other possible formation channels for ultramassive WDs, most importantly binary evolution, specifically the merger of double-degenerate systems \citep{Segretain97,Marsh97,Liebert05,GarciaBerro12}.

A handful of ultramassive WDs ($\geq 1.2$ \msun) have been found in nature. \citet{VK08} have reviewed the evidence for ultramassive WDs and find much of it compelling. However, since we cannot see below the photosphere of these WDs, our understanding of their interiors is essentially superficial.

Direct evidence that ultramassive WDs harbor ONe cores comes from heavy isotope anomalies found in classical novae, which match predicted abundances from explosive nucleosynthesis on massive WDs with ONeMg cores \citep{Gehrz98}. Additionally, two oxygen-rich WDs were recently discovered in the Sloan Digital Sky Survey (SDSS; \citealt{Gansicke10}). These WDs are likely exposed ONe cores, as the observed O/C abundance ratio indicates a very low overall carbon mass fraction, a prediction for some of the most massive progenitors avoiding core collapse \citep{Iben97}.

A more direct test of ultramassive WD core composition would be to find a massive WD undergoing pulsations. Asteroseismology offers the unique opportunity to use these pulsations to probe below the photosphere and into the interior of stars, and has had numerous successful applications with WDs (see reviews by \citealt{WinKep08,FontBrass08,Althaus10}). We have thus engaged in a search for pulsations in massive WDs in or near the DAV (or ZZ Ceti) instability strip, a region for which WDs with hydrogen-dominated atmospheres have the appropriate temperature to develop a hydrogen partial ionization zone, which in turn drives global pulsations. That search has already yielded multiple new massive DAVs \citep{Kepler12,Castanheira13} after the 1.1~\msun\ BPM~37093 discovered by \citet{Kanaan92}.

Here we report a new success in that search: the discovery of the most massive pulsating WD known, GD~518. Model fits to the optical spectrum first reported by \citet{Gianninas11} show this is a \teff\ $= 12{,}030 \pm 210$ K WD with \logg\ $=9.08 \pm 0.06$, which would correspond to a mass of $1.20 \pm 0.03$ \msun\ using the ONe WD models of \citet{Althaus05} or a mass of $1.23 \pm 0.02$ \msun\ using the CO WD models of \citet{Wood95}. 

In this Letter we present our discovery of pulsations in GD~518. In Sections~\ref{sec:observations} and \ref{sec:analysis} we outline our observations and analysis. We conclude with a discussion of the impact of this finding in Section~\ref{sec:discussion}.



\section{Observations}
\label{sec:observations}

We targeted GD~518 (WD~J165915.11+661033.3) as a candidate ultramassive pulsating WD based on model atmosphere fits to its optical spectrum. The object was first classified in \citet{Gianninas11}. Evolutionary models by \citet{Althaus07} suggest that such a WD has a cooling age of roughly 1.7 Gyr and an absolute $V$-band magnitude of 13.6 mag, which indicates that GD~518 ($g$=17.2 mag) is roughly 53 pc from Earth.

\begin{figure}[t]
\centering{\includegraphics[width=0.55\columnwidth]{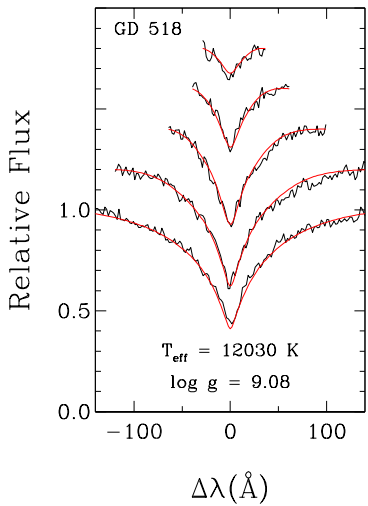}}
\caption{The individual Balmer line profiles (black) of GD 518. The lines range from H$\beta$ (bottom) to H8 (top), each offset by a factor of 0.2 in relative flux. The model fits (red), first reported by \citet{Gianninas11}, derive the atmosphere parameters and show this is a high-surface-gravity WD, with \logg\ $=9.08 \pm 0.06$. This corresponds to a mass of $1.20 \pm 0.03$ \msun. \label{fig:GD518spec}}
\end{figure}

We display the optical spectrum analyzed by \citet{Gianninas11} in Figure~\ref{fig:GD518spec}. The spectrum was obtained in 2009~March using the 2.3 m telescope at Steward Observatory, equipped with the Boller \& Chivens spectrograph at a resolution of $\sim$6~\AA\ full width at half-maximum (FWHM), covering a wavelength range from roughly $3700-5200$ \AA. A more detailed explanation of the observations, models, and fitting can be found in Sections~$2-4$ of \citet{Gianninas11}.

We obtained additional spectroscopy from the FLWO 1.5 m telescope in 2013~April using the FAST spectrograph \citep{Fabricant98}. These observations, four 20 minute exposures at a resolution of 1.7~\AA\ FWHM, cover a wavelength range from $3600-5500$ \AA. Using the same models and fitting method as in \citet{Gianninas11}, we confirm that this WD has a very high surface gravity. Although this new summed spectrum is much lower signal-to-noise ratio (S/N$\sim$15), our fits formally yield \teff\ $= 12{,}100 \pm 370$ K and \logg\ $=9.00 \pm 0.09$, which agree with the previous determination within the stated uncertainties. We thus adopt the primary parameters derived from the higher quality (S/N$\sim$55) spectrum analyzed in \citet{Gianninas11}, displayed in Figure~\ref{fig:GD518spec}.

A $12{,}030$ K temperature puts GD~518 inside an extrapolated empirical instability strip for DAVs of high mass \citep{Castanheira13}. In fact, \citet{Gianninas11} noted that GD~518 was a ``most intriguing'' candidate to target for possible pulsations.

We obtained time-series photometric observations of GD~518 at the McDonald Observatory over nine nights in 2013~March, eight of them consecutive, and five nights in 2013~April, for a total of more than 42.9 hr of coverage over 33 nights. A full journal of observations can be found in Table~\ref{tab:jour}. We used the Argos instrument, a frame-transfer CCD mounted at the prime focus of the 2.1m Otto Struve telescope \citep{Nather04}, to obtain $5-10$ s exposures on this $g=17.2$ mag WD. The seeing averaged 2.0\arcsec\ and conditions were generally fair. Observations were obtained through a 3mm BG40 filter to reduce sky noise.

\begin{deluxetable}{lcccc}
\tabletypesize{\scriptsize}
\tablecolumns{5}
\tablewidth{0.375\textwidth}
\tablecaption{Journal of Photometric Observations. \label{tab:jour}}
\tablehead{
\colhead{UT Date} & \colhead{Length} & \colhead{Seeing} & \colhead{Exp.} & \colhead{\ensuremath{P_{\rm{+}}}[\ensuremath{A_{\rm{+}}}]}
\\ \colhead{} & \colhead{(hr)} & \colhead{(\arcsec)} & \colhead{(s)} & \colhead{(s)[(mma)]} }
\startdata
2013 Mar 10 & 3.1 & 3.9 & 10 & 437.6[4.9] \\
2013 Mar 12 & 3.0 & 2.2 & 5 & 437.6[2.0] \\
2013 Mar 13 & 3.0 & 1.7 & 5 & 418.4[1.8] \\
2013 Mar 14 & 2.3 & 1.4 & 5 & 438.0[4.3] \\
2013 Mar 15 & 2.5 & 1.7 & 5 & 441.0[3.8] \\
2013 Mar 16 & 3.0 & 1.8 & 5 & 441.4[6.5] \\
2013 Mar 17 & 3.4 & 1.9 & 5 & 439.9[6.2] \\
2013 Mar 18 & 3.5 & 2.2 & 5 & 438.5[4.6] \\
2013 Mar 19 & 2.9 & 1.5 & 5 & 437.0[4.1] \\
2013 Apr 4 & 3.6 & 1.8 & 5 & 440.0[5.4] \\
2013 Apr 6 & 4.4 & 2.0 & 5 & 441.1[3.7] \\
2013 Apr 7 & 3.1 & 1.4 & 5 & 524.2[1.6] \\
2013 Apr 9 & 2.6 & 2.5 & 5 & 519.1[3.6] \\
2013 Apr 12 & 2.9 & 1.7 & 5 & 514.2[4.3]
\enddata
\end{deluxetable}

The raw science frames were calibrated by dark subtraction and flat-fielding. We performed weighted aperture photometry on the calibrated frames using the external IRAF package $\textit{ccd\_hsp}$ written by Antonio Kanaan (the reduction method is outlined in \citealt{Kanaan02}). We divided the sky-subtracted light curves by the sum of the three nearest brighter comparison stars in the field to correct for transparency variations, and applied a timing correction to each observation to account for the motion of the Earth around the barycenter of the solar system \citep{Stumpff80,Thompson09}.

\begin{figure}[t]
\centering{\includegraphics[width=0.9\columnwidth]{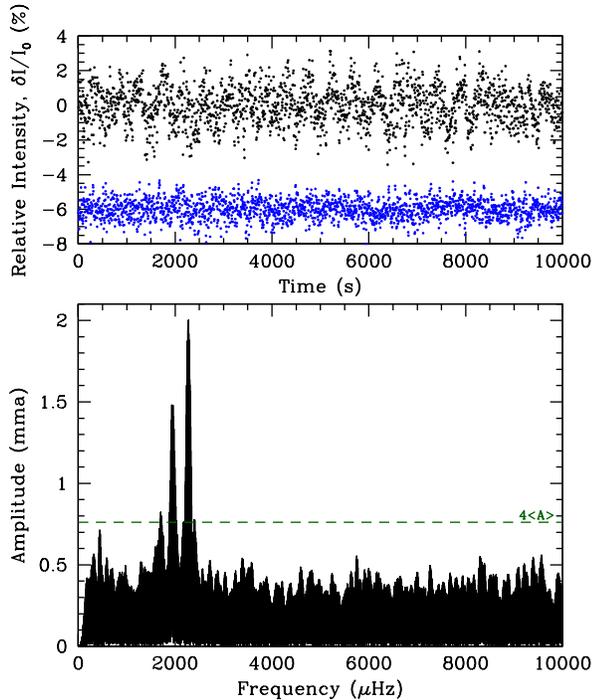}}
\caption{The top panel shows high-speed photometry of GD~518, this a portion from 2013 March 16. The brightest comparison star is shown in blue, offset by $-6$\%. For both we have co-added the data by two points, slightly smoothing the light curve. The bottom panel shows a Fourier transform of our entire data set to date, some $29{,}985$ points taken during more than 42.9 hr of observations in 2013 March and April. We mark the 4$\langle {\rm A}\rangle$ reference, described in the text, as a dashed green line. \label{fig:GD518lc}}
\end{figure}

The top panel of Figure~\ref{fig:GD518lc} shows the light curve of GD~518 from 2013 March 16. A two-frequency solution to this 3.0-hr run finds variability at $441.36 \pm 0.66$ s ($6.5 \pm 0.4$ mma\footnote[1]{1 mma = 0.1\% relative amplitude}) and $514.1 \pm 2.4$ s ($2.5 \pm 0.4$ mma).

The bottom panel of Figure~\ref{fig:GD518lc} shows a Fourier transform (FT) for our entire data set, some $29{,}985$ points from more than 42.9 hr of observations in 2013 March and April. We display the 4$\langle {\rm A}\rangle$ reference line, calculated from the average amplitude, $\langle {\rm A}\rangle$, of the FT of the entire data set from 0 to $10{,}000$ \muhz. 



\section{Light Curve Analysis}
\label{sec:analysis}

The optical light curve of GD~518 shows low-amplitude but statistically significant variability at multiple periods, ranging from roughly $425-595$ s, with amplitudes that can reach up to 0.7\% over a single night of observations. This can be seen by eye in the top panel of Figure~\ref{fig:GD518lc}, as well as in the FT of our entire data set in the bottom panel of that same figure.

We have attempted to identify the periodicities present in the star, which will form the basis for future asteroseismic modeling. Complicating our analysis, however, is the fact that the amplitudes (and perhaps frequencies) of the observed variability are not consistent from night-to-night. In fact, the FT for a few nights had no significant peaks above 1.5 mma. We have included the period (\ensuremath{P_{\rm{+}}}) and amplitude (\ensuremath{A_{\rm{+}}}) of the highest peaks for each night in Table~\ref{tab:jour}.

There is thus some strong frequency and/or amplitude modulation occurring in GD~518 acting on the timescale of days, perhaps caused by beating of closely spaced periodicities or perhaps due to a physical mechanism in the star. We have therefore broken up the data into different subsets of minimum length allowed by the frequency splitting in the overall frequency solution: the first five nights (2013 Mar $10-14$), the second five nights (2013 Mar $15-19$), and the final nine nights (2013 Apr $4-12$).

We present a frequency solution for each subset in Table~\ref{tab:GD518freq}. It was determined by computing an FT, then a nonlinear least-squares fit on the frequency with the highest amplitude, then prewhitening by that frequency until there are no peaks above a 4$\langle {\rm A}\rangle$ significance line, which came from the average amplitude of an FT from 0 to $10{,}000$ \muhz\ of the unprewhitened data. We have included the 508 s periodicity in the solution for our first subset even though it is not above 4$\langle {\rm A}\rangle$, based on its presence in other subsets. For more realistic estimates, the quoted uncertainties in Table~\ref{tab:GD518freq} are not formal least-squares uncertainties to the data but rather the product of 1000 Monte Carlo simulations of perturbed data using the software package Period04 \citep{Lenz05}.

\begin{deluxetable}{lccc}
\tablecolumns{4}
\tablewidth{0.45\textwidth}
\tablecaption{Frequency solution for GD 518
  \label{tab:GD518freq}}
\tablehead{\colhead{ID} & \colhead{Period} & \colhead{Frequency} & \colhead{Amplitude}
\\ \colhead{} & \colhead{(s)} & \colhead{($\mu$Hz)} & \colhead{(mma)} }
\startdata
\multicolumn{4}{c}{\bf Overall Frequency Solution} \\
$f_{1}$	&	440.2	$\pm$	1.5	&	2271.7	$\pm$	7.6	& \\
$f_{2}$	&	513.2	$\pm$	2.4	&	1948.6	$\pm$	9.2	& \\
$f_{3}$	&	583.7	$\pm$	1.5	&	1713.3	$\pm$	4.5	& \\
\multicolumn{4}{c}{\bf Using First Five Nights (Mar 10$-$14)} \\
$f_{1a}$	&	438.47	$\pm$	0.64	&	2280.7	$\pm$	3.3	&	2.92	$\pm$	0.44		\\
$f_{1b}$	&	438.098	$\pm$	0.057	&	2282.59	$\pm$	0.30	&	2.24	$\pm$	0.48\\
$f_2$	&	508.2	$\pm$	1.5	&	1967.9	$\pm$	5.9	&	1.30	$\pm$	0.29		\\
\multicolumn{4}{c}{\bf Using Second Five Nights (Mar 15$-$19)} \\
$f_{1a}$	&	439.6	$\pm$	4.5	&	2275	$\pm$	24	&	4.05	$\pm$	0.57	 	\\
$f_{1b}$	&	438.89	$\pm$	0.16	&	2278.45	$\pm$	0.82	&	2.57	$\pm$	0.24 \\
$f_{1c}$	&	440.26	$\pm$	0.25	&	2271.4	$\pm$	1.3	&	2.42	$\pm$	0.39	 \\
$f_{2a}$	&	511.3	$\pm$	2.9	&	1956	$\pm$	11	&	2.0	$\pm$	1.3 \\
$f_{2b}$	&	509.405	$\pm$	0.099	&	1963.08	$\pm$	0.38	&	1.8	$\pm$	1.6	\\
$f_3$	&	518.99	$\pm$	0.14	&	1926.82	$\pm$	0.52	&	1.49	$\pm$	0.37	 \\
$f_4$	&	592	$\pm$	33	&	1690	$\pm$	95	&	1.24	$\pm$	0.49	 \\
\multicolumn{4}{c}{\bf Using Last Nine Nights (Apr 4$-$12)} \\
$f_1$	&	519.238	$\pm$	0.043	&	1925.90	$\pm$	0.16	&	2.51	$\pm$	0.48	 \\
$f_{2a}$	&	441.244	$\pm$	0.046	&	2266.32	$\pm$	0.23	&	2.38	$\pm$	0.41	 \\
$f_{2b}$	&	440.156	$\pm$	0.062	&	2271.92	$\pm$	0.32	&	2.12	$\pm$	0.40	 \\
$f_3$	&	512.6	$\pm$	5.3	&	1951	$\pm$	20	&	1.63	$\pm$	0.47	 \\
\multicolumn{4}{c}{\bf Using All Data (Mar 10$-$Apr 12)} \\
$f_{1a}$	&	442.12	$\pm$	0.42	&	2261.8	$\pm$	2.1	&	2.38	$\pm$	0.73	 \\
$f_{1b}$	&	441.15	$\pm$	0.17	&	2266.81	$\pm$	0.88	&	2.36	$\pm$	0.72	 \\
$f_{1c}$	&	439.5	$\pm$	1.4	&	2275.5	$\pm$	7.3	&	1.94	$\pm$	0.53	 \\
$f_2$	&	519.2	$\pm$	1.8	&	1925.9	$\pm$	6.7	&	1.55	$\pm$	0.59	 \\
$f_{1d}$	&	440.59	$\pm$	0.47	&	2269.7	$\pm$	2.4	&	1.21	$\pm$	0.52	 \\
$f_{3a}$	&	511.455	$\pm$	0.009	&	1955.207	$\pm$	0.035	&	1.21	$\pm$	0.26 \\
$f_{3b}$	&	510.824	$\pm$	0.008	&	1957.622	$\pm$	0.032	&	1.13	$\pm$	0.18	 \\
$f_{1e}$	&	437.79	$\pm$	0.13	&	2284.20	$\pm$	0.67	&	1.13	$\pm$	0.46 \\
$f_{4a}$	&	503.800	$\pm$	0.010	&	1984.914	$\pm$	0.040	&	0.95	$\pm$	0.23 \\
$f_{4b}$	&	501.44	$\pm$	0.50	&	1994.3	$\pm$	2.0	&	0.93	$\pm$	0.24 \\
$f_5$	&	426.71	$\pm$	0.86	&	2343.5	$\pm$	4.7	&	0.81	$\pm$	0.34	 \\
$f_6$	&	587.25	$\pm$	0.96	&	1702.8	$\pm$	2.8	&	0.78	$\pm$	0.24	
\enddata
\end{deluxetable}

\begin{figure}[t]
\centering{\includegraphics[width=0.85\columnwidth]{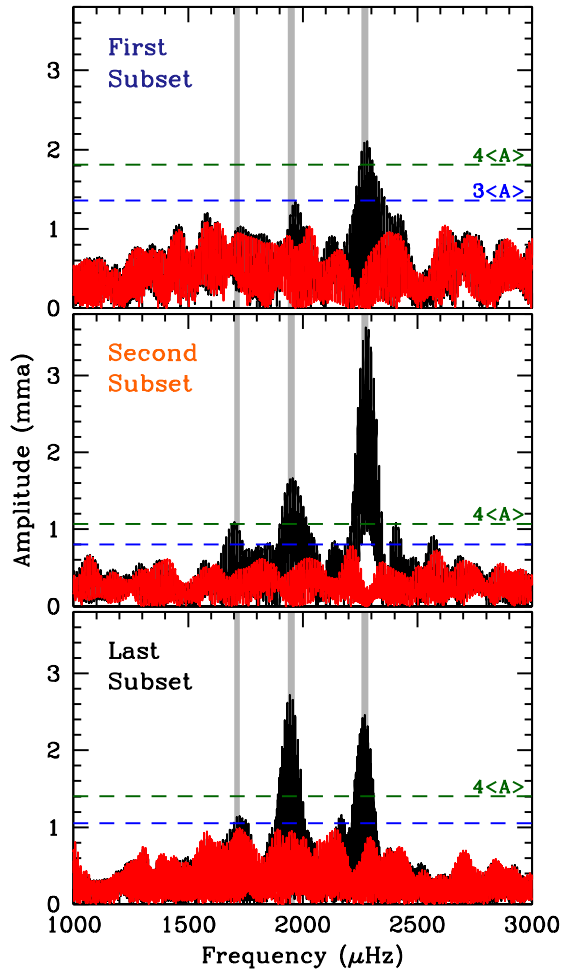}}
\caption{Fourier transforms, in black, of the light curves of our first five nights of data (top panel), our second five nights of data (middle panel) and our last nine nights of data (bottom panel). The frequency solutions for each subset are described in Table~\ref{tab:GD518freq}. In each case we also display in red the Fourier transform of the residuals after pre-whitening by the significant frequencies. We mark the 4$\langle {\rm A}\rangle$ and 3$\langle {\rm A}\rangle$ significance lines as dashed green and blue lines, respectively. The vertical gray lines show the 1$\sigma$ uncertainties for the overall frequency solution adopted in Table~\ref{tab:GD518freq}. \label{fig:halves}}
\end{figure}

We calculate the chance these detections are real by computing the false alarm probability (FAP) using the formalism described in \citet{Kepler93}. We find that all periodicities in each subset have a FAP $> 99.9$\% except for $f_2$ in the first subset, which has a FAP of $91.0$\%.

Computing a full frequency solution for our entire dataset, using the same method as we have for each subset, yields 12 formally significant frequencies, many of which are quite closely spaced (see the bottom panel of Table~\ref{tab:GD518freq}). Since we observe large-scale amplitude changes over the course of days, we have chosen not to adopt these 12 frequencies as a formal solution, because we cannot confirm the coherence of each periodicity. Some frequencies may represent sampling artifacts or frequency drifting rather than truly excited modes in the star. Still, we include these 12 frequencies in Table~\ref{tab:GD518freq} since our full dataset allows us to detect low-amplitude features that may be additional independent periods. Every periodicity in the frequency solution for the entire dataset has a FAP $> 99.9$\% except for $f_6$, which has a 99.8\% FAP.

We calculate a more conservative overall frequency solution by fitting a Lorentzian function to the three significant bands of power for the FT of each of the three subsets, shown in Figure~\ref{fig:halves}. We adopt the mean of the centroids, weighted by their FWHM, as the overall frequency solution in Table~\ref{tab:GD518freq}, with the uncertainty determined by the standard deviation of the three measurements.

Figure~\ref{fig:halves} shows the Fourier analysis for each subset. We display the original FT for that subset in black, overlaid with the FT prewhitened by the frequencies marked as significant in Table~\ref{tab:GD518freq}. It is evident that the amplitudes of the variability near 440 s and 513 s (2270 and 1949 \muhz, respectively) change significantly, even over the timescale of a few days.



\section{Discussion and Conclusions}
\label{sec:discussion}

We have discovered pulsations in GD~518, which is to date the most massive pulsating WD known. This star has a mass of roughly 1.2 \msun, derived from model fits to its pressure-broadened Balmer lines. The object offers the best opportunity, to date, to explore the interior of a possible ONe-core WD using asteroseismology.

Since our best current evidence on the high-mass nature of GD~518 rests on its optical spectrum, we have been careful to ensure that this WD truly has high surface gravity. Masses of WDs derived from the spectroscopic method, as we have here with GD~518, show an unphysical upturn in derived surface gravity for effective temperatures below $11{,}500$ K (e.g., \citealt{Koester09}). However, the models used to calculate the surface gravity of GD~518 include improved Stark broadening profiles with non-ideal gas effects, which have slightly moderated this upturn \citep{Tremblay11,Kleinman13}. A full three-dimensional treatment of convection for a WD atmosphere near $12{,}000$ K and \logg\ $=9.0$ shows that corrections to the one-dimensional models we have used for this spectroscopic analysis do not diverge by more than $0.1-0.15$ dex (P.-E.~Tremblay 2013, private communication). Additionally, we do not observe evidence for splitting of the Balmer lines caused by a high surface magnetic field, which can sometimes be confused as a high-surface-gravity WD \citep{Kepler13}.

Fits to follow-up spectroscopy on GD~518 agree with the high surface gravity first reported in \citet{Gianninas11}, which confirms a high-mass interpretation for this WD. Additionally, the star is in the footprint of the SDSS, and matching $ugriz$ colors with synthetic models\footnote[2]{http://www.astro.umontreal.ca/$_{\widetilde{~}}$bergeron/CoolingModels}
suggests this is an ultramassive WD \citep{HB06,KS06,Tremblay11}. Obtaining a parallax distance to GD~518 will help settle its mass.


There is theoretical support to expect that it will be possible to distinguish the core composition of a massive WD. \citet{Corsico04} explored the adiabatic pulsational properties of massive WDs and found several noticeable differences between CO-core models and ONe-core models of a $1.05$ \msun\ WD, the only mass they calculated. Their ONe-core models were characterized by strong deviations in their forward period spacing, and the mean period spacing for their ONe models was noticeably larger than the mean period spacing for their CO-core modes. Additionally, the pulsation modes in their ONe-core models had consistently lower kinetic energies than those in the CO-core models.

However, the reason they found lower kinetic energies (and larger period spacings) for pulsations in their ONe-core models is that those ONe-core models were significantly more crystallized ($>90$\% by mass) at the same temperature, $11{,}810$ K, than their CO-core models ($\sim$50\% by mass). Crystallization occurs when the Coulomb energy between neighboring ions becomes more than two orders of magnitude larger than the thermal energy of the ions in the WD core, and is a naturally occurring stage of WDs as they cool \citep{Salpeter61,DAntona90}. A 1.2 \msun\ WD should be significantly more crystallized at a similar temperature than a 1.05 \msun\ WD. We expect that pulsation energy would be largely excluded from the interior crystallized mass.

With less of the stellar material participating in the global pulsations, it is conceivable that the oscillations have less mode inertia, and can vary on shorter timescales relative to the pulsation periods. Indeed, we observe large amplitude changes in this massive WD (see Figure~\ref{fig:halves}), which may be a consequence of its large crystallized mass fraction. This relatively short-term amplitude modulation, especially in which pulsation amplitudes fall below detectability, has been seen before in other massive pulsating WDs, notably BPM~37093 and SDSS J005047.60-002316.9 \citep{Kanaan05,Castanheira10}. As the highest-mass pulsating WD ever discovered, GD~518 will provide rich insight into the physics of crystallization, as initiated by studies of BPM~37093 \citep{Metcalfe04}.

However, the degeneracy in parameters caused by crystallization will pose a significant challenge to finding a robust asteroseismic differentiation of the core composition of GD~518. We expect to need a significant number of observed independent pulsation modes in order to overcome the many free parameters in our asteroseismic fits. Still, we are encouraged by the number of independent periods we have already determined (at least three) with a relatively short, single-site campaign.

It is possible that the three highest-amplitude periods we observe --- at $440.2\pm1.5$~s, $513.2\pm2.4$~s, and $583.7\pm1.5$~s --- are of the same spherical degree ($\ell$), since they are spaced by roughly 73.0 s and 70.5 s, respectively. They are unlikely consecutive radial orders. The 90\% crystallized 1.15 \msun\ CO-core models of \citet{MW99} find mean period spacings for $\ell=2$ modes of $15-25$ s, depending on the hydrogen layer mass. Likewise, \citet{Corsico04} expect period spacings of roughly 20 s for $\ell=2$ modes of a 1.05 \msun\ ONe-core model. Period spacings for $\ell=1$ modes should be $\sqrt{3}$ times longer.

Aside from the imprint the core chemical profile makes on the pulsation spectrum, the core composition also affects the rate of cooling for a WD. This is a consequence of the fact that the cooling time of a WD is inversely proportional to the mean atomic weight of the ions in the core. In some cases we can directly measure this cooling rate by monitoring, long-term, the rate of period change of stable pulsation modes in a DAV (e.g., \citealt{Kepler05,Mukadam13}). Measuring the rate of period change of any coherent modes will allow another direct test of core composition for this ultramassive WD, albeit a longer-term endeavor.

Successfully unraveling both the overall mass and the core composition of GD~518 will constrain intermediate-mass stellar evolution. It also provides an opportunity to put an upper limit on the primary in a Type Ia supernovae progenitor system, which theory predicts is a CO-core WD, since an ONeMg-core WD is expected to collapse due to electron capture before detonation as a Type Ia supernova \citep{Nomoto84,Nomoto87}.


\acknowledgments

We thank R.~E. Falcon and E.~L. Robinson for useful comments. This work is supported by the Norman Hackerman Advanced Research Program, under grant 003658-0252-2009, and by the National Science Foundation, under grant AST-0909107. We acknowledge the McDonald Observatory staff for their support, especially John Kuehne and Dave Doss.

\end{document}